\documentclass[twocolumn,showpacs,preprintnumbers,amsmath,amssymb]{revtex4}
\usepackage{epsf}
\usepackage{graphicx}% Include figure files
\usepackage{dcolumn}% Align table columns on decimal point
\usepackage{bm}% bold math

\begin{document}

\title{Inflationary Attractor from Tachyonic Matter}
 
\author{Zong-Kuan Guo}
 \email{guozk@itp.ac.cn} 
\author{Yun-Song Piao}
 \email{yspiao@itp.ac.cn} 
\author{Rong-Gen Cai}
 \affiliation{{Institute of Theoretical Physics, Chinese Academy of
 Sciences, P.O. Box 2735, Beijing 100080, China.}}
\author{Yuan-Zhong Zhang}
 \affiliation{{CCAST (World Lab.), P.O. Box 8730, Beijing 100080, China.\\
 Institute of Theoretical Physics, Chinese Academy of
 Sciences, P.O. Box 2735, Beijing 100080, China.}}

\date{\today}% It is always \today, today,

\begin{abstract}
We study the complete evolution of a flat and homogeneous universe
dominated by tachyonic matter. We demonstrate the attractor behaviour of
the tachyonic inflation using the Hamilton-Jacobi formalism. We else obtain
analytical approximations to the trajectories of the tachyon field in
different regions. The numerical calculation shows that an initial
non-vanishing momentum does not prevent the onset of inflation. The
slow-rolling solution is an attractor. 
\end{abstract}

\maketitle

%%========================section 1 =============================
\section{Introduction}

The study of non-BPS objects such as non-BPS branes, brane-antibrane
configurations and spacelike branes has recently attracted great attention
given its implications for string/M-theory and cosmology.
The tachyon field associated with unstable D-branes, might be responsible
for cosmological inflation at early epochs due to tachyon condensation near
the top of the effective potential~\cite{DDD}, and could contribute to
some new form of cosmological dark matter at late times~\cite{SEN}.
Several authors have investigated the process of rolling of the tachyon
in the cosmological background~\cite{GWG,GS}.
In the slow roll limit in FRW cosmology, the 
exact solution of tachyonic inflation with exponential potential is
found~\cite{MPT}.

A question which has not yet been addressed in the literature on tachyonic
inflation is the issue of constraints on the phase space of initial
conditions for inflation which arise when one takes into account the fact
that in the context of cosmology the momenta of the tachyon field
cannot be neglected in the early universe. For models of the type of
chaotic inflation, the work of~\cite{HAF} shows that most of the
energetically accessible field value space give rise to a sufficiently
long period of slow roll inflation. However, for models of the type of
new inflation, allowing for non-vanishing initial field momenta may
dramatically reduce the phase space of initial conditions for which
successful inflation results, and the attractor is the slow rolling
solution~\cite{DSG}.

In this paper we investigate the constrains on the initial conditions
of inflation with tachyon rolling down an exponential potential
in phase space required for successful inflation. 
We demonstrate the attractor behaviour of the tachyonic
inflation using the Hamilton-Jacobi formalism.
We else use an explicitly numerical
computation of the phase space trajectories and obtain analytical
approximations to the trajectories of the tachyon in different regions.
We find that in phase space there exists a curve that attracts most of 
the solutions. 

%========================section 2===============================
\section{Tachyonic Matter Cosmology}

According to Sen~\cite{SEN}, the effective action of the tachyon field in
the Born-Infeld form can be written as
\begin{equation}
\label{E1}
S=\int d^4x\sqrt{-g}\left(\frac{1}{2\kappa ^2}R
+V(T)\sqrt{1+g^{\mu \nu}\partial _\mu T\partial _\nu T}\right)
\end{equation} 
where T is the tachyon field minimally coupled to gravity.
The rolling tachyon in a spatially
flat FRW cosmological model can be described by a fluid with a positive
energy density $\rho$ and a negative pressure $P$ given by
\begin{eqnarray}
\rho &=& \frac{V(T)}{\sqrt{1-\dot{T}^2}}, \\
P &=& -V(T)\sqrt{1-\dot{T}^2},
\end{eqnarray}
Thus
\begin{equation}
\omega=\frac{P}{\rho}=-(1-\dot{T}^2)
\end{equation}
Note that $-1\le \omega \le 0$, and a universe dominated by this rolling
tachyonic matter will smoothly evolve from a phase of accelerating 
expansion to a phase dominated by a non-relativistic fluid~\cite{GWG}.
The evolution equation of the tachyon field minimally coupled to gravity,
and the Friedmann equation are
\begin{eqnarray}
\label{4T}
\frac{\dot{\Pi}}{1-\Pi ^2}+3H\Pi+\frac{V'(T)}{V(T)}=0, \\
H^2=\frac{\kappa ^2}{3}\frac{V(T)}{\sqrt{1-\Pi ^2}},
\label{5F}
\end{eqnarray}
where $\Pi \equiv \dot{T}$ denotes the velocity of tachyon.
A universe dominated by
tachyon field would go under accelerating expansion as long as
$\Pi^2<2/3$ which is very different from the condition of inflation
for non-tachyonic field, $\dot{\phi}^2<V(\phi)$. The tachyon potential
$V(T)\to 0$ as $T\to \infty$, but its exact form is not known at
present~\cite{AAGS}.
Sen has argued that the qualitative dynamics of string theory tachyons
can be described by (\ref{E1}) with the exponential potential~\cite{SENN}
\begin{equation}
V(T)=V_0e^{-\alpha T},
\end{equation}
where $\alpha$ is the tachyon mass. The cosmological aspects of rolling
tachyon with exponential potential are investigated~\cite{MPT}. 
In what follows we will consider
(\ref{E1}) with exponential potential in purely phenomenological context
without claiming any identification of $T$ with the string tachyon field.

%%========================section 3===========================
\section{Inflationary Attractor}

The Hamilton-Jacobi formulation~\cite{SB}
is a powerful way of rewriting the equations
of motion, which allows an easier derivation of many inflation results. We
concentrate here on the homogeneous situation as applied to spatially flat
cosmologies, and demonstrate the attractor behaviour of the tachyonic
inflation using the Hamilton-Jacobi formalism~\cite{LPB}.

Differentiating Eq.(\ref{5F}) with respect to $t$ and substituting in
Eq.(\ref{4T}) gives
\begin{equation}
\dot{T}=-\frac{2}{3}\frac{H^\prime (T)}{H^2(T)}
\end{equation}
where primes denote derivatives with respect to the tachyon field $T$, which
gives the relation between $T$ and $t$. This allows us to write the
Friedmann equation in the first-order form
\begin{equation}
\label{HJ}
[H^\prime (T)]^2-\frac{9}{4}H^4(T)=-\frac{\kappa ^4}{4}V^2(T)
\end{equation}
Eq.(\ref{HJ}) is the Hamilton-Jacobi equation. It allows us to
consider $H(T)$, rather than $V(T)$, as the fundamental quantity to be
specified. Suppose $H_0(T)$ is any solution to Eq.(\ref{HJ}), which can be
either inflationary or non-inflationary. Add to this a linear homogeneous
perturbation $\delta H(T)$; the attractor condition will be satisfied if it
becomes small as $T$ increases. Substituting $H(T)=H_0(T)+\delta H(T)$ into
Eq.(\ref{HJ}) and linearizing, we find that the perturbation obeys
\begin{equation}
H_0^\prime \delta H^\prime \simeq \frac{9}{2}H_0^3\delta H
\end{equation}
which has the general solution
\begin{equation}
\delta H(T)=\delta H(T_i)\exp \left(\frac{9}{2}
\int_{T_i}^T\frac{H_0^3(T)}{H_0^\prime (T)}dT\right)
\end{equation}
where $\delta H(T_i)$ is the value at some initial point $T_i$. Because
$H^\prime _0$ and $dT$ have opposing signs, the integrand within the
exponential term is negative definite, and hence all linear perturbations
do indeed die away. If there is an inflationary solution, all linear
perturbations approach it at least exponentially fast as the tachyon field
rolls.

%%========================section 4=============================
\section{Phase Portrait and Cosmological Evolution}

We choose different initial conditions $T$, in the range $T\ge 0$, and
$\Pi$ in the range $0\le \Pi \le1$ and we obtain the phase portrait in
the $(T,\Pi)$ plane. Figure 1 shows that there exists a curve that attracts
most of the trajectories, in the $(x,y)$ plane where $x=\alpha T$ and
$y=\Pi$ are dimensionless coordinates. The initial kinetic term decays
rapidly and does not prevent the onset of inflation.  

\begin{figure}[h!]
\centering\leavevmode\epsfysize=6.6cm \epsfbox{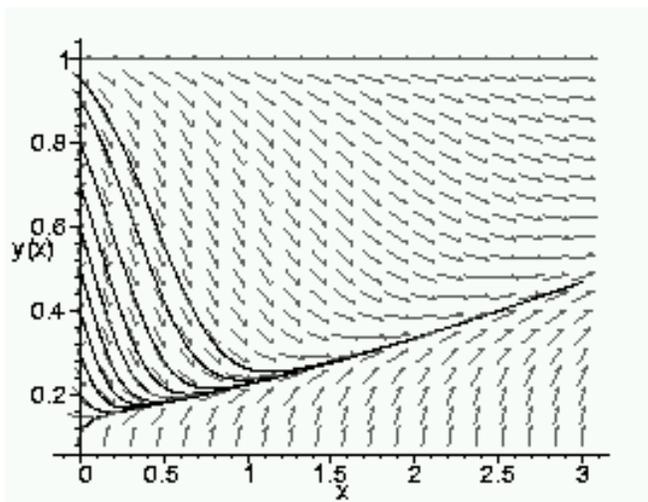}

\

\caption{Phase portrait for tachyonic cosmology.} 
\end{figure}

The behavior of the trajectories can be also analyzed analytically. To
understand the evolution of the tachyon field we define two regions
$P$ and $K$ in the $(T,\Pi)$ plane as indicated in Figure 2. The region
$P$ is the region where the potential dominates over the energy density,
and the region $K$ is the region where the kinetic energy dominates.

\begin{figure}[h!]
\centering\leavevmode\epsfysize=5.7cm \epsfbox{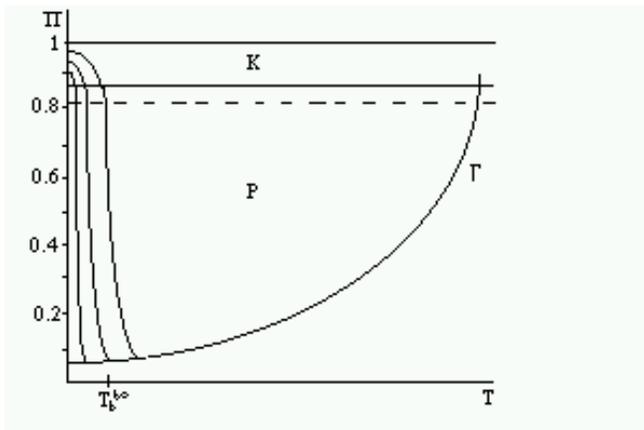}

\

\caption{Sketch of the different regions in the ($T,\Pi$) phase plane.} 
\end{figure}

{\bf Curve $\Gamma$.}
This curve describes the slow rolling solution where the evolution Eq.
(\ref{4T}) and the Friedman Eq.(\ref{5F}) in expanding universe can
be approximated by 
\begin{eqnarray}
3H\Pi +\frac{V'(T)}{V(T)}=0, \\
H^2=\frac{\kappa ^2}{3}V(T),
\end{eqnarray}
from which it follows that
\begin{eqnarray}
\Pi &=&\frac{\alpha}{3\beta}e^{\frac{\alpha }{2}T} \\
T&=&\frac{2}{\alpha}\ln{\left(\frac{3\beta}{\alpha}\Pi \right)}
\end{eqnarray}
where $\beta \equiv \sqrt{\kappa ^2V_0/3}$.
The expression for the number $N$ of e-foldings of inflation can be written
as
\begin{equation}
\label{EN}
N=\ln{\frac{a_e}{a_b}}=\kappa ^2\int_{T_e}^{T_b}\frac{V^2}{V''}dT
=\left.\frac{\kappa ^2}{\alpha^2}V(T)\right|_{T_e}^{T_b},
\end{equation}
where $T_b$ is the value of the field at the point where it reaches curve
$\Gamma$ and $T_e$ is the value of the field at the end of the slow rolling
phase. In order to obtain enough e-foldings of slow roll inflation
the value of $T_b$ for such a trajectory must satisfy
\begin{equation}
\label{TB}
0\le T_b<T_b^{60}
\end{equation}
where $T_b^{60}$ is calculated from Eq.(\ref{EN}) with $N=60$.

{\bf Region $P$.}
In this region, the potential dominates over the energy density. The 
potential force $V'(T)$ is negligible compared to the friction term since
the friction coefficient is proportional to the potential. The evolution
Eq.(\ref{4T}) and Friedman Eq.(\ref{5F}) are approximately  
\begin{eqnarray}
\dot{\Pi}+3H\Pi=0 \\
H^2=\frac{\kappa ^2}{3}V(T)
\end{eqnarray}
so that 
\begin{eqnarray}
\Pi &=&\Pi_ {pk}-\frac{6\beta}{\alpha}\left(e^{-\frac{\alpha}{2}T_{pk}}
-e^{-\frac{\alpha}{2}T}\right) \\
T&=&-\frac{2}{\alpha}\ln{\left[e^{-\frac{\alpha}{2}T_{pk}}-
\frac{\alpha}{6\beta}\left(\Pi _{pk}-\Pi \right)\right]}
\end{eqnarray}
where $T_{pk}$ and $\Pi _{pk}$ are the values at the boundary between region
$P$ and region $K$. Let us now denote by $T_b$ and $\Pi _b$ the values
of the inflaton and its momentum when the trajectory reaches the
slow roll curve.  
\begin{equation}
T_b=-\frac{2}{\alpha}\ln{\left[e^{-\frac{\alpha}{2}T_{pk}}
-\frac{\alpha}{6\beta}\Pi _{pk}\right]}
\end{equation}
where we have neglected $\Pi _b$ since it is exponentially smaller than
$\Pi _{pk}$. In general the unconventional
forms of the tachyonic energy density and pressure make the cosmology
with tachyon field differ from that with a normal scalar field, and make
it difficult to separate kinetic term from potential term. We assume that
$V(T)$ and $V(T)(\frac{1}{\sqrt{(1-\dot{T}^2)}}-1)$
are regarded as potential and kinetic term respectively. Therefore, 
the boundary value between region $P$ and region $S$ is
$\Pi _{pk}=\frac{\sqrt{3}}{2}$. So
\begin{equation}
T_b=-\frac{2}{\alpha}\ln{\left[e^{-\frac{\alpha}{2}T_{pk}}
-\frac{\alpha}{4\sqrt{3}\beta}\right]}
\end{equation}
To lead to sufficient inflation, from (\ref{TB}) such initial conditions
must satisfy
\begin{equation}
-\frac{2}{\alpha}\ln{\left[e^{-\frac{\alpha}{2}T_{pk}}
-\frac{\alpha}{4\sqrt{3}\beta}\right]}<T_b^{60} 
\end{equation}

{\bf Region $K$.}
It is the region of kinetic energy domination where $V'(T)$ is negligible 
compared to the friction term and (\ref{4T}) and (\ref{5F}) become
\begin{eqnarray}
\frac{\dot{\Pi }}{1-\Pi ^2}+3H\Pi =0 \\
H^2=\frac{\kappa ^2}{3}\frac{V(T)}{\sqrt{1-\Pi ^2}}
\end{eqnarray}
These can be integrated and we find: 
\begin{eqnarray}
\Pi _2F_1\left(\frac{1}{2},\frac{3}{4},\frac{3}{2},\Pi ^2\right)&=&
\Pi _{pk2}F_1\left(\frac{1}{2},\frac{3}{4},
\frac{3}{2},\Pi _{pk}^2\right) \nonumber \\
&&-\left.\frac{6\beta}{\alpha}e^{-\frac{\alpha}{2}T}\right|_{T}^{T_{pk}}
\end{eqnarray}
\begin{eqnarray}
T=-\frac{2}{\alpha}\ln{\left[e^{-\frac{\alpha}{2}T_{pk}}-\left.
\frac{\alpha}{6\beta}\Pi _2F_1\left(\frac{1}{2},\frac{3}{4},
\frac{3}{2},\Pi ^2\right)\right|_{\Pi}^{\Pi _{pk}}\right]}
\end{eqnarray}
where $_2F_1\left(\frac{1}{2},\frac{3}{4},\frac{3}{2},\Pi ^2\right)$
is Gauss hypergeometric function.

%%=======================section 5==============================
\section{Conclusions and Discussions}

To demonstrate the attractor behaviour of the tachyonic inflation, We
use the Hamilton-Jacobi formalism, which greatly simplifies the analysis.
Adding to any solution to Eq.(\ref{HJ}) a linear homogeneous perturbation,
we find the perturbation die away exponentially. The attractor behaviour
indicates that, regardless of initial conditions, the late-time solutions
are the same up to a time shift, which cannot be measured~\cite{LMS}.
We else use an explicitly 
numerical computation of the phase space trajectories and obtain analytical
approximations to the trajectories of the tachyon in different regions.
One can easily verify from Figure 1 and Figure 2 that these approximations
are in very good agreement with the numerical results, and that the slow
rolling solution is the late-time attractor. Although the initial kinetic
term decays rapidly and does not prevent the onset of inflation, allowing
for non-vanishing initial field momenta around $T=0$ may dramatically
reduce the phase space of initial conditions for which successful inflation
results.

According to the picture of tachyonic inflation, the homogeneous
tachyon field near the top of the potential rolls down towards the
minimum of the potential at $T\to \infty$. Tachyonic matter behaves at
late time as a pressureless gas of massive particles.
In most versions of the theory of reheating,
production of particles occurs only when the inflaton field oscillates
near the minimum of its effective potential. However, the effective
potential of the rolling tachyon does not have any minimum at finite $T$,
so this mechanism does not work. It is unclear
how the universe could be reheated in the framework of tachyon cosmology.
Recently, some studies pointed out that as the tachyon evolves into
the late-time, the coupling to the closed string becomes more and more
large~\cite{GHY}. These results motivate us to expect that the tachyon could
emit closed string radiation~\cite{MGAS,OY}, such as graviton and dilation,
into the bulk and eventually settles in the finite minimum.

\begin{acknowledgments}
It is a pleasure to acknowledge helpful discussions with G.N.Felder and
R.S.Tung. This project was in part supported by NNSFC under Grant
Nos. 10175070 and 10047004 as well as also by NKBRSF G19990754.
\end{acknowledgments}

\end{document}